\documentclass[prb,twocolumn,amsmath,amssymb,aps,superscriptaddress,citeautoscript,longbibliography]{revtex4-2}

\usepackage{feynmp}
\usepackage{amsmath}
\usepackage{graphicx}
\usepackage{multirow}
\usepackage{latexsym}
\usepackage{textcomp}
\usepackage{verbatim}
\usepackage{color}
\usepackage{bm}
\usepackage{subfigure}
\usepackage{everysel}
\usepackage{keyval}
\usepackage{ragged2e}
\usepackage{dsfont}
\usepackage{amssymb}
\usepackage{enumitem}
\usepackage{pstricks}
\usepackage{mathtools}
\usepackage{kotex} % Use texlive 2021

\usepackage{braket}
\usepackage{slashed} 
\usepackage{hyperref}        
\hypersetup{
     colorlinks = true,
     citecolor = red,    
     linkcolor = cyan,
     }

\usepackage{graphicx}
\usepackage{xcolor}
\usepackage{amsmath}
\usepackage{bbold}
\usepackage{amsfonts}
\usepackage{bbm}
\usepackage{comment}
\usepackage{lettrine}
\usepackage{orcidlink}

\newcommand{\osum}{{%
    \setbox0\hbox{\circ}%
    \rlap{\hbox to \wd0{\hss\sum\hss}}\box0
}}

\begin{document}

\title{Intrinsic orbital Hall effect in a nonuniform electric field}

\author{Min Ju Park\,\orcidlink{0000-0003-1451-0469}}
\thanks{Contact author: mp7@ualberta.ca}
\affiliation{Department of Physics \& Center for Quantum Dynamics of Angular Momentum, Pohang University of Science and Technology, Pohang 37673, Korea}
\affiliation{Department of Physics, University of Alberta, Edmonton, Alberta T6G 2E1, Canada}

\author{Jongjun M. Lee\,\orcidlink{0000-0002-9786-1901}}
\thanks{Contact author: jongjun@ualberta.ca}
\affiliation{Department of Physics \& Center for Quantum Dynamics of Angular Momentum, Pohang University of Science and Technology, Pohang 37673, Korea}
\affiliation{Department of Physics, University of Alberta, Edmonton, Alberta T6G 2E1, Canada}
\affiliation{Quantum Horizons Alberta, University of Alberta, Edmonton, Alberta T6G 2E1, Canada}

\author{Hyun-Woo Lee\,\orcidlink{0000-0002-1648-8093}}
\thanks{Contact author: hwl@postech.ac.kr}
\affiliation{Department of Physics \& Center for Quantum Dynamics of Angular Momentum, Pohang University of Science and Technology, Pohang 37673, Korea}

\begin{abstract}
Geometric analysis of electronic Bloch states offers a universal framework for understanding electronic properties, yet its role in the transport of orbital angular momentum remains unexplored. In this work, we establish an analytic connection between orbital angular momentum transport and the geometric properties of Bloch wave functions in electronic systems. Focusing on the intrinsic orbital Hall effect in the dc limit under a spatially nonuniform electric field, we show that its conductivity can be expressed in terms of universal geometric quantities, such as the orbital Berry curvature and quantum metric. This formulation provides a term-by-term correspondence with the geometric description of intrinsic charge Hall transport established in previous studies. Using a tight-binding model, we further illustrate that the higher-order orbital Hall response can exhibit enhanced sensitivity to the orientation of an anisotropic sample. Our work deepens the understanding of diverse intrinsic transverse transport phenomena and the role of quantum geometry in electronic systems.
\end{abstract}

\date{\today}
\maketitle

\section{Introduction}
Spin-orbitronics studies the transport of both electron spin and orbital angular momentum (OAM), extending the field of spintronics~\cite{go2021orbitronics,jo2024spintronics,lee2024electric}. While the OAM is typically quenched in the matter at equilibrium, it can be transported or exhibit a net value in non-equilibrium conditions~\cite{kittel2018introduction,go2018intrinsic}. Recent experimental and theoretical studies have provided strong evidence that the transport of OAM in metals can be even more efficient than that of spin angular momentum~\cite{jo2018gigantic,hayashi2023observation,go2023long,choi2023observation,moriya2024observation}. Various related phenomena have been proposed, including the orbital Hall effect (OHE)~\cite{bernevig2005orbitronics,tanaka2008intrinsic,pezo2022orbital}, orbital Edelstein effect~\cite{johansson2021spin,el2023observation,lee2024orbital}, and unusual relaxation dynamics~\cite{sohn2024dyakonov,kabanov2024impact}. The field is expanding to encompass other quasi-particles, such as phonons and magnons~\cite{zhang2014angular,neumann2020orbital,park2020phonon,go2024magnon}. To support comprehensive studies, a range of experimental techniques has been introduced, including orbital torque in heterostructures~\cite{lee2021orbital,dutta2022observation,fukunaga2023orbital}, magneto-optical Kerr effects~\cite{choi2023observation,lyalin2023magneto}, magnetoresistance~\cite{ding2022observation,sala2023orbital,hayashi2023orbital}, and pump-probe measurements~\cite{chen2019role,tauchert2022polarized}.

The Bloch wave function of an electron in a periodic lattice possesses a geometric structure in reciprocal space~\cite{kittel2018introduction}. This structure provides a universal framework for describing the electron's dynamics, independent of the specific details of the system in condensed matter. For example, the Berry curvature and quantum metric are directly related to the linear response~\cite{haldane2004berry,sinova2004universal,xiao2005berry,torma2018quantum,torma2022superconductivity}, while other geometric quantities are associated with higher-order optical responses, such as the photovoltaic effect~\cite{ahn2022riemannian}. Additionally, a few studies have proposed that the dc current of charge or spin is connected to various geometric quantities under a nonuniform electromagnetic field~\cite{oscar2023multipole,zhong2016gyrotropic,kozii2021intrinsic,zhang2022geometric}. A previous study has established a link between the quantum metric and orbital magnetic susceptibility~\cite{piechon2016geometric}. However, the connection between the transport of OAM and the geometry of the Bloch wave function has not yet been explored much~\cite{pezo2023orbital,liu2023covariant,lee2025universal,urru2025optical}.

\begin{figure}[t!]
    \centering
    \includegraphics[width=0.9\linewidth]{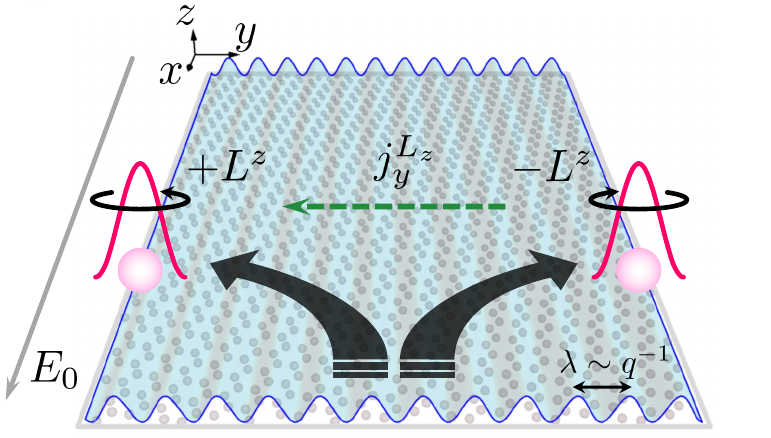}
    \caption{Schematic illustration of the intrinsic orbital Hall effect in a nonuniform electric field. The electric field $\hat{x}E_{0}e^{i(qy-\omega t)}$ is applied such that its spatial variation (blue wave pattern on the plane) is perpendicular to its overall direction (gray arrow). The resulting current of electronic OAM $j^{L_z}_{y}$ flows transversely (green dashed arrow). The pink curve represents an electronic wave packet that exhibits self-rotation along the direction indicated by the black arrow, corresponding to the OAM $\pm L^{z}$.}
    \label{fig1}
\end{figure}

In this study, we elucidate the link between the transport of OAM and the geometric characteristics of Bloch wave functions in electronic systems. We explore the intrinsic OHE in the dc limit under a spatially nonuniform electric field~\cite{oscar2023multipole,zhong2016gyrotropic,kozii2021intrinsic,zhang2022geometric}. We demonstrate that the electronic orbital Hall response can also be characterized by universal quantities, such as the orbital Berry curvature and quantum metric, analogous to previous research on the intrinsic charge Hall transport~\cite{kozii2021intrinsic}. Furthermore, by drawing an analogy to charge transport, we derive the explicit form of the higher-rank geometric quantities relevant to orbital responses, which incorporates the OAM operator. We further illustrate, using a tight-binding model, that this higher-order response tends to be more sensitive to the sample orientation than the conventional zeroth-order response. Our results are general, since they require only a real two-orbital system respecting both inversion and time-reversal symmetries, and therefore do not rely on a specific lattice model. This framework is applicable to various two-orbital systems with nontrivial orbital textures, such as $p_x$ and $p_y$ orbital systems on square and triangular lattices~\cite{luttinger1956quantum,han2023microscopic}.

This paper is organized as follows. In Sec.~\ref{sec_model}, we introduce the general two-band model and specify symmetries. In Sec.~\ref{sec_geometry}, we define the relevant quantum geometric quantities. In Sec.~\ref{sec_response}, we formulate the orbital Hall response to a spatially nonuniform electric field. In Sec.~\ref{sec_result}, we express the orbital Hall response in terms of geometric quantities. In Sec.~\ref{sec_discussion}, we discuss the correspondence between our result and the charge Hall conductivity, together with its implications. In Sec.~\ref{sec_numerics}, we provide a numerical analysis using a tight-binding model and illustrate the orientation dependence of the higher-order response. Finally, Sec.~\ref{Sec_conclusion} summarizes our findings.

\section{Model}\label{sec_model}
%{\color{blue}Two-band system. Describe the model with clarity and completeness without redundancy.}
We consider a two-dimensional electronic system described within a two-orbital subspace spanned by orthonormal real orbitals on a lattice, such as $p_x$ and $p_y$ orbitals, as schematically illustrated in Fig.~\ref{fig1}. We assume that the system preserves both time-reversal and spatial inversion symmetries, and neglect spin-orbit coupling and electron-electron interactions~\cite{wu2008p,tokatly2010orbital}. Under these conditions, the OAM (and magnetization) in each eigenstate is suppressed, and the system is effectively described by a two-band free-electron model~\cite{han2023microscopic}. 

The Hamiltonian density in momentum space is then written as
\begin{equation}
    \mathcal{H}_{\bf k}|n_{\bf k}\rangle = \epsilon_{n,\mathbf{k}} |n_{\bf k}\rangle,
\end{equation}
where $n=1,2$ denotes the band index, $\epsilon_{n,\mathbf{k}}$ is the corresponding energy, and $\mathbf{k}$ is the crystal momentum. Here, $\lvert n_{\mathbf{k}} \rangle$ denotes the Bloch eigenstate of the Hamiltonian at momentum $\mathbf{k}$ in the $n$-th band. By symmetry, the Hamiltonian can be represented as a real matrix, and its eigenstates can be chosen to be real~\cite{ahn2019stiefel,ahn2019failure}. We note that, while we assume the symmetries, we do not specify any particular lattice geometry or model parameters.

For the OAM operator $\hat{L}_{z}$ along the $z$ direction, the Bloch eigenstates in our system satisfy
\begin{equation}
\langle n_{\bf k}| \hat{L}_{z} |m_{\bf k}\rangle = i l_{0} \epsilon_{nm},
\label{Loperator}
\end{equation}
where $n,m=1,2$, $\epsilon_{nm}$ denotes the two-dimensional Levi-Civita symbol, and $l_{0}$ is a real constant determined by the details of the system. For example, in the $p_x$-$p_y$ model, one has $l_{0}=1$. Further details are provided in Appendix~\ref{Appx_OAM_Operator}.

\section{Quantum geometry}\label{sec_geometry}
Since our result of the intrinsic response is closely related to quantum geometry, we briefly review and define quantum geometric quantities, such as the Berry connection and the quantum metric~\cite{provost1980riemannian,xiao2010berry}.

The real non-Abelian Berry connection, a matrix-valued generalization of the Berry connection, is defined as~\cite{ahn2019failure}
\begin{equation}
    A^{j}_{mn,\mathbf{k}} \equiv \langle m_{\bf k} | \partial_{k_{j}} | n_{\bf k} \rangle.
\label{A}
\end{equation}
The diagonal (Abelian) components correspond to the intraband contributions; however, they vanish under the assumed symmetries. In addition, for the off-diagonal terms, $A^{j}_{mn,\mathbf{k}}=-A^{j}_{nm,\mathbf{k}}$, therefore we write $A^{j}_{12,\mathbf{k}}=A^{j}_{\bf k}$ without the band index.

The (Abelian) quantum geometric tensor for the $n$-th band is defined as 
\begin{equation}
    Q^{ij}_{n,\mathbf{k}} = \langle \partial_{k_{i}} n_{\bf k}| \Pi_{n,\mathbf{k}} | \partial_{k_{j}}n_{\bf k}\rangle,
\end{equation}
where $\Pi_{n,\mathbf{k}}=1-|n_{\mathbf{k}}\rangle\langle n_{\mathbf{k}}|$ denotes the complementary projector~\cite{kozii2021intrinsic}. Its real and imaginary parts correspond to the quantum metric ($g^{ij}_{n,\mathbf{k}}$) and the Berry curvature ($\Omega^{ij}_{n,\mathbf{k}}$), respectively. Under the symmetries of our model, the latter vanishes. Furthermore, since our model consists of only two bands, the quantum metric can be expressed in terms of the non-Abelian Berry connection as
\begin{equation}
    g^{ij}_{n,\mathbf{k}} = A_{n\bar{n},\mathbf{k}}^{i}A_{n\bar{n},\mathbf{k}}^{j}
\end{equation}
where $\bar{n}\neq n$. This tensor is symmetric under the exchange of $i$ and $j$. Moreover, the band index $n$ is redundant in our model, since $g^{ij}_{n,\mathbf{k}} = g^{ij}_{\bar{n},\mathbf{k}}$. We therefore write $g^{ij}_{\mathbf{k}}$.

Finally, we define the quantum geometric connection for three directions $i,j,l$ as
\begin{equation}
    c^{ijl}_{n,\mathbf{k}} = 
    \langle n_{\bf k} | (\partial_{k_{i}} \partial_{k_{j}}  \Pi_{n,\mathbf{k}})(\partial_{k_{l}}  \Pi_{n,\mathbf{k}}) | n_{\bf k} \rangle.
\end{equation}
In our two-band system, one can show that
\begin{equation}
c^{ijl}_{n,\mathbf{k}} = A^{l}_{\mathbf{k}}\partial_{k_i} A^{j}_{\mathbf{k}},
\end{equation}
which is real-valued and independent of the band index $n$. We therefore omit the band index and write $c^{ijl}_{\mathbf{k}}$. The symmetrized combination
\begin{equation}
    T^{ijl}_{\mathbf{k}} = 
    \frac{1}{3}\,\mathrm{Im}\!\left(
        c^{ijl}_{\bf k} + c^{jli}_{\bf k} + c^{lij}_{\bf k}
    \right)
\label{Eq_Tijl_1}
\end{equation}
is analogous to a Christoffel symbol of the quantum metric and is gauge invariant~\cite{ahn2020low}; however, it vanishes in our system due to symmetry.

\section{Formalism for a nonuniform electric field}\label{sec_response}
We study the electronic response to a spatially varying electric field. Specifically, we consider an electric field oriented along the $x$ direction with a finite wave vector $q$ along the $y$ direction, as illustrated in Fig.~\ref{fig1}. The field is written as
\begin{equation}
    \mathbf{E}(\mathbf{r}, t) = \hat{\mathbf{x}} E_0 e^{i (q y - \omega t)}+\text{c.c.},
\label{Eq_E_1}
\end{equation}
where $\hat{\mathbf{x}}$ is the unit vector along the $x$ direction. Here, $q$ denotes the wave vector characterizing the spatial modulation of the nonuniform electric field, rather than the momentum of a propagating photon. Under this electric field, Bloch states with different momenta become mixed and collectively contribute to the response, as observed in the charge and spin conductivities~\cite{kozii2021intrinsic,zhang2022geometric}.

We define the OAM current operator corresponding to OAM polarized along the $i$ direction and flowing along the $j$ direction as
\begin{equation}
   \hat{j}^{L_i}_{j,\mathbf{k}} = \frac{1}{2} (\hat{L}_{i}\hat{v}^{j}_{\mathbf{k}}+\hat{v}^{j}_{\mathbf{k}}\hat{L}_{i}),
\end{equation}
where $\hat{L}_{i}$ denotes the OAM operator along the $i$ direction and $\hat{v}^{j}_{\mathbf{k}}=\partial_{k_j}H(\mathbf{k})$ is the velocity operator along the $j$ direction. Here and throughout this paper, we set $\hbar=1$. Since we consider atomic orbitals localized at each lattice site, $\hat{L}_{i}$ is independent of the crystal momentum $\mathbf{k}$~\cite{bernevig2005orbitronics,go2018intrinsic,go2021orbitronics}.

Under the spatially varying electric field in Eq.~\eqref{Eq_E_1}, we consider the OHE for OAM polarized along the $z$ axis and flowing along the $y$ direction. The corresponding OAM current is given by
\begin{equation}
j^{L_{z}}_{y}(\mathbf{r},t)=
%\sum_{q,\omega}
\sigma^{L_{z}}_{yx}(q,\omega)E_{0}e^{i(qy-\omega t)}+\text{c.c}.
\end{equation}
We focus on the static (dc) limit ($\omega \rightarrow 0$), where the orbital Hall conductivity is defined as $\sigma^{L_{z}}_{yx}(q)= \lim_{\omega \rightarrow 0} \sigma^{L_{z}}_{yx}(q,\omega)$. In addition, we consider the long-wavelength regime of the field modulation, corresponding to small $q$. Accordingly, we expand the orbital Hall conductivity as
\begin{equation}
    \sigma^{L_{z}}_{yx}(q)
    = \sum^{\infty}_{n=0} q^{n} \sigma^{L_{z},(n)}_{yx}.
\end{equation}
Here, the odd-order terms in $q$ vanish due to spatial inversion symmetry.

The orbital Hall conductivity is evaluated using the Kubo-Greenwood formula as follows~\cite{kozii2021intrinsic,zhang2022geometric}.
\begin{widetext}
\begin{equation}
    \sigma^{L_{z}}_{yx}(q) = 
    -\frac{e}{S} 
    \sum_{n,m,\bm{k}} (f_{n,\mathbf{k}-\mathbf{q}/2}-f_{m,\mathbf{k}+\mathbf{q}/2})
    \frac{
    \mathrm{Im}\!\left[
        \langle n_{\bm{k}-\bm{q}/2}| \hat{j}^{L_z}_{y,\mathbf{k}} |m_{\bm{k}+\bm{q}/2}\rangle 
        \langle m_{\bm{k}+\bm{q}/2}| \hat{v}^{x}_{\mathbf{k}} |n_{\bm{k}-\bm{q}/2}\rangle
    \right]
    }{
    \left(\epsilon_{n,\bm{k}-\bm{q}/2}-\epsilon_{m,\bm{k}+\bm{q}/2}\right)^{2}
    },
\label{Eq_OHE_1}
\end{equation}
\end{widetext}
where $f_{n,\mathbf{k}}$ is the Fermi-Dirac distribution, $e$ is the electron charge, $S$ is the area of the system, $m$ and $n$ denote band indices, and $\mathbf{q}=q\hat{y}$; see Appendix~\ref{Appx_Kubo_1} for the detailed derivation~\cite{marder2010condensed}. In the following calculation, Eq.~\eqref{Eq_OHE_1} is evaluated by expanding the Bloch states for a slowly varying field, namely in the small-$q$ limit, as
\begin{equation}
|n_{\mathbf{k}\pm\mathbf{q}/2}\rangle\simeq|n\rangle\pm \frac{q}{2}|\partial_y n\rangle+\frac{q^2}{8}|\partial_y^2 n\rangle.
\end{equation}

\section{Orbital Hall conductivity}
\label{sec_result}
We calculate the orbital Hall conductivity in response to a spatially modulated electric field using Eq.~\eqref{Eq_OHE_1}. Due to spatial inversion symmetry, only even powers of $q$ contribute to the conductivity. Intraband processes generate only odd powers of $q$ and therefore do not contribute in our system. Accordingly, we focus on the even-order ($q^{2n}$) terms arising from interband processes.

We first consider the zeroth-order term in $q$, which corresponds to the conventional OHE. The result is given by
\begin{equation}
\sigma^{L_{z},(0)}_{yx} = 
\frac{e}{S} 
\sum_{\mathbf{k}} 
(f_{1,\mathbf{k}}-f_{2,\mathbf{k}}) 
\frac{ \mathrm{Im}[L^{z}_{\mathbf{k}}]
\bar{v}^{y}_{\mathbf{k}} 
A^{x}_{\mathbf{k}}}{ \delta\epsilon_{\bf k} },
\label{mysigma1}
\end{equation}
where $\delta\epsilon_{\bf k} =\epsilon_{1,\mathbf{k}}-\epsilon_{2,\mathbf{k}}$ is the energy difference between the bands, $L^{z}_{\mathbf{k}} = \langle 1_{\mathbf{k}} | \hat{L}_z | 2_{\mathbf{k}} \rangle$ is the off-diagonal component of the OAM operator, $\bar{v}^{i}_{\mathbf{k}} = (v^{i}_{11,\mathbf{k}} + v^{i}_{22,\mathbf{k}})/2$ is the symmetrized velocity, and $v^{i}_{mn,\mathbf{k}}=\langle m_{\mathbf{k}}| \hat{v}^{i}_{\mathbf{k}}|n_{\mathbf{k}}\rangle$. 

We note that the zeroth-order term in Eq.~\eqref{mysigma1} coincides with the well-known result for the OHE under a uniform electric field. In particular, it can be written as a sum over the orbital Berry curvature, $\sigma^{L_{z},(0)}_{yx} =  -\frac{e}{S}\sum_{\mathbf{k}} (f_{1,\mathbf{k}}-f_{2,\mathbf{k}})\,\Omega^{L_{z}}_{yx,\mathbf{k}}$, where the orbital Berry curvature $\Omega^{L_{z}}_{yx,\mathbf{k}}$ is defined via the Kubo formula~\cite{go2018intrinsic} as 
\begin{equation}
\Omega^{L_{z}}_{yx,\mathbf{k}} = 
\mathrm{Im}\Big[
\frac{ \langle 1_{\mathbf{k}} | \hat{j}^{L_z}_{y,\mathbf{k}} | 2_{\mathbf{k}} \rangle \langle 2_{\mathbf{k}} | \hat{v}^{x}_{\mathbf{k}} | 1_{\mathbf{k}} \rangle }{ \delta\epsilon_{\bf k}^{2} }
\Big].
\label{Eq_OBC_1}
\end{equation}
This expression reproduces Eq.~\eqref{mysigma1}, demonstrating that our result is consistent with previous studies.

Next, we examine the second-order term in $q$, which provides the leading correction to the conventional OHE. The result is given by
\begin{equation}
\sigma^{L_{z},(2)}_{yx} =  
-\frac{e}{S} 
\sum_{\mathbf{k}} (f_{1,\mathbf{k}}-f_{2,\mathbf{k}}) 
\sum_{j=1}^{3}\sigma^{L_{z},(2)}_{yx,\mathbf{k},j},
\end{equation}
where
\begin{equation}
\sigma^{L_{z},(2)}_{yx,\mathbf{k},1} =   
\frac{\mathrm{Im}[L^{z}_{\mathbf{k}}]\bar{v}^{y}_{\mathbf{k}}}{ 2\delta\epsilon_{\bf k} }
A^{x}_{\mathbf{k}} (A^{y}_{\mathbf{k}})^{2},
\label{Eq_s21}
\end{equation}
\begin{equation}
\begin{aligned}
\sigma^{L_{z},(2)}_{yx,\mathbf{k},2} =& 
\frac{\mathrm{Im}[L^{z}_{\mathbf{k}}]\bar{v}^{y}_{\mathbf{k}}}{ 8\delta\epsilon_{\bf k}^{2} }
( -2 A^{x}_{\mathbf{k}} \partial_{k_{y}}\delta v^{y}_{\bf k} 
+ \delta v^{x}_{\bf k} \partial_{k_{y}}A^{y}_{\mathbf{k}} ),
\end{aligned}
\label{Eq_s22}
\end{equation}
and
\begin{equation}
\begin{aligned}
\sigma^{L_{z},(2)}_{yx,\mathbf{k},3} =& 
\frac{\mathrm{Im}[L^{z}_{\mathbf{k}}]\bar{v}^{y}_{\mathbf{k}}}{ 4\delta\epsilon_{\bf k}^{3} }
( -3 (\delta v^{y}_{\bf k})^{2}A^{x}_{\mathbf{k}} 
+ \bar{v}^{x}_{\mathbf{k}} \delta v^{y}_{\bf k} A^{y}_{\mathbf{k}}  ).
\end{aligned}
\label{Eq_s23}
\end{equation}
Here, $\delta v^{i}_{\bf k} = v^{i}_{11,\mathbf{k}} - v^{i}_{22,\mathbf{k}}$. The result is expressed in terms of the non-Abelian Berry connection, velocity matrix elements, and band energies. As in the zeroth-order case, these expressions can be reorganized into a geometric form involving the orbital Berry curvature and its higher-order counterparts.

We rewrite the second-order term of the orbital Hall conductivity in terms of geometric quantities such as the quantum metric, as well as their extended geometric counterparts (hereafter referred to as extended geometric quantities), such as the orbital Berry curvature [Eq.~\eqref{Eq_OBC_1}]. This is done in a manner analogous to the zeroth-order case. The result is given by
\begin{widetext}
\begin{equation}
\begin{aligned}
\sigma^{L_{z},(2)}_{yx}  =& \frac{e}{2S} 
\sum_{\mathbf{k}} (f_{1,\mathbf{k}}-f_{2,\mathbf{k}}) 
\Big[
g^{yy}_{\mathbf{k}}  \Omega^{L_{z}}_{yx,\mathbf{k}}
- \frac{1}{\delta \epsilon_{\mathbf{k}}}
\Big(
\frac{\delta v^{y}_{\mathbf{k}}}{2} 
\partial_{k_{y}} \Omega^{L_{z}}_{yx,\mathbf{k}}
+ \frac{3\delta v^{y}_{\mathbf{k}}}{2} 
T^{L_{z}}_{yyx,\mathbf{k}}
- \frac{\delta v^{x}_{\mathbf{k}}}{4} 
T^{L_{z}}_{yyy,\mathbf{k}}
\Big)
+ 4 \frac{ \bar{v}^{y}_{\bf k}\delta v^{y}_{\bf k} }
{\delta \epsilon_{\mathbf{k}}^{2}} 
\, \Omega^{L_{z}}_{yx,\mathbf{k}}
\Big],
\end{aligned}
\label{Eq_Result_ours_1}
\end{equation}
\end{widetext}
where the first, second, and third terms in the square brackets correspond to Eqs.~\eqref{Eq_s21}, \eqref{Eq_s22}, and \eqref{Eq_s23}, respectively. The following quantities are defined for the second part,
\begin{equation}
\begin{aligned}
T^{L_{z}}_{yyx,\mathbf{k}} =& - \frac{A^{x}_{\bf k}\partial_{k_{y}}\delta\epsilon_{\bf k}}{\delta\epsilon^{2}_{\bf k}} \text{Im}[\langle 1_{\bf k}| \hat{j}^{L_{z}}_{y,\mathbf{k}} |2_{\bf k} \rangle], \\
T^{L_{z}}_{yyy,\mathbf{k}} =& -\frac{\delta\epsilon_{\bf k}\partial_{k_{y}}A^{y}_{\bf k}}{\delta\epsilon^{2}_{\bf k}} \text{Im}[\langle 1_{\bf k}| \hat{j}^{L_{z}}_{y,\mathbf{k}} |2_{\bf k} \rangle].
\end{aligned}
\label{Eq_T_yyx}
\end{equation}
The second-order contribution expressed in terms of the extended geometric quantities has a form nearly identical to that of the corresponding contribution for the charge Hall effect, written in terms of geometric quantities, reported in previous studies~\cite{kozii2021intrinsic}. This constitutes our main result.

\begin{table*}[t!]
\begin{tabular}{lcc}
\hline \hline \noalign{\vskip 4pt}
 & Charge & Orbital angular momentum \\[4pt] \hline \noalign{\vskip 4pt}
Current & $\begin{aligned}\hat{j}^{y}_{\mathbf{k}}=e \hat{v}^{y}_{\bf k}\end{aligned}$ & $\begin{aligned}\hat{j}^{L_{z}}_{y,\mathbf{k}}=\frac{1}{2}\{ \hat{L}_{z},\hat{v}^{y}_{\bf k} \}\end{aligned}$ \\[6pt] \hline \noalign{\vskip 4pt}
Conductivity & $\begin{aligned} &\sigma_{yx}(q)=\sum_{n}q^{n}\sigma^{(n)}_{yx} \\ &\sigma^{(2)}_{yx}= \frac{e^{2}}{2S}\sum_{\mathbf{k},j}\tilde{f}_{\bf k}\sigma^{(2)}_{yx,\mathbf{k},j} \end{aligned} $  & $\begin{aligned} &\sigma^{L_{z}}_{yx}(q)=\sum_{n}q^{n}\sigma^{L_{z},(n)}_{yx}\\ &\sigma^{L_{z},(2)}_{yx}= -\frac{2e}{S}\sum_{\mathbf{k},j}\tilde{f}_{\bf k}\sigma^{L_{z},(2)}_{yx,\mathbf{k},j} \end{aligned}$ \\[22pt] \hline \noalign{\vskip 4pt}
$\begin{aligned} &\text{1st part} \\ &(\sigma^{(2)}_{yx,\mathbf{k},1},\: \sigma^{L_{z},(2)}_{yx,\mathbf{k},1}) \end{aligned}$ & $\begin{aligned}g^{yy}_{\mathbf{k}}\Omega^{yx}_{\mathbf{k}}\end{aligned}$ & $\begin{aligned} g^{yy}_{\mathbf{k}}\Omega^{L_{z}}_{yx,\mathbf{k}}\end{aligned}$ \\[12pt] \hline \noalign{\vskip 4pt}
$\begin{aligned}&\text{2nd part}\\ &(\sigma^{(2)}_{yx,\mathbf{k},2},\: \sigma^{L_{z},(2)}_{yx,\mathbf{k},2}) \end{aligned}$ & $\begin{aligned}-\frac{1}{\delta \epsilon_{\mathbf{k}}}\Big( \frac{\delta v^{y}_{\bf k}}{3}\partial_{k_{y}}\Omega^{yx}_{\mathbf{k}} + \frac{\delta v^{y}_{\bf k}}{2}T^{yyx}_{\mathbf{k}} - \frac{\delta v^{x}_{\bf k}}{2}T^{yyy}_{\mathbf{k}} \Big)\end{aligned}$  & $\begin{aligned}-\frac{1}{\delta \epsilon_{\mathbf{k}}}\Big( \frac{\delta v^{y}_{\bf k}}{2}\partial_{k_{y}}\Omega^{L_{z}}_{yx,\mathbf{k}} + \frac{3\delta v^{y}_{\bf k}}{4} T^{L_{z}}_{yyx,\mathbf{k}} - \frac{\delta v^{x}_{\bf k}}{4} T^{L_{z}}_{yyy,\mathbf{k}} \Big)\end{aligned}$ \\[12pt] \hline \noalign{\vskip 4pt}
$\begin{aligned}&\text{3rd part}\\ &(\sigma^{(2)}_{yx,\mathbf{k},3},\: \sigma^{L_{z},(2)}_{yx,\mathbf{k},3}) \end{aligned}$ & $\begin{aligned}- \frac{2v^{y}_{11,\mathbf{k}}v^{y}_{22,\mathbf{k}}}{\delta \epsilon_{\mathbf{k}}^{2}} \Omega^{yx}_{\mathbf{k}}\end{aligned}$ & $\begin{aligned}4 \frac{ \bar{v}^{y}_{\bf k}\delta v^{y}_{\bf k} }{\delta \epsilon_{\mathbf{k}}^{2}}\Omega^{L_{z}}_{yx,\mathbf{k}}\end{aligned}$ \\[12pt] \hline
\end{tabular}
\caption{Summary of the orbital Hall conductivity under a nonuniform electric field, compared with the intrinsic charge Hall conductivity reported in Ref.~\cite{kozii2021intrinsic}. Here, $\tilde{f}_{\mathbf{k}} = f_{1,\mathbf{k}} - f_{2,\mathbf{k}}$.}
\label{Table_summary_1}
\end{table*}

\section{Correspondence with charge Hall conductivity} \label{sec_discussion}
We discuss the correspondence between our orbital Hall conductivity in Eq.~\eqref{Eq_Result_ours_1} and the charge Hall conductivity in a two-band electronic system reported in a previous study~\cite{kozii2021intrinsic}. Overall, both results share a similar structure, consisting of three contributions. A direct comparison between the corresponding terms is summarized in Table~\ref{Table_summary_1}.

Among the terms in Eq.~\eqref{Eq_Result_ours_1}, the first two and the last one correspond directly to those in the charge conductivity reported in the previous study, upon replacing the Berry curvature $\Omega_{yx,\mathbf{k}}$ with $\Omega^{L_z}_{yx,\mathbf{k}}$. This correspondence is natural, since the zeroth-order contributions to the orbital and charge Hall conductivities are given by the sums of the orbital Berry curvature and the Berry curvature, respectively~\cite{go2018intrinsic}. 

Notably, the quantum-metric contribution $g^{yy}_{\mathbf{k}}\Omega^{L_z}_{yx,\mathbf{k}}$ has the same structure as in the charge Hall response, with the Berry curvature replaced by the orbital Berry curvature. Following the wave-packet interpretation of Ref.~\cite{kozii2021intrinsic}, $g^{yy}_{\mathbf{k}}$ represents the spread of the corresponding maximally localized Wannier orbital along the field-gradient direction~\cite{marzari1997maximally}, which is independent of whether the transported quantity is charge or OAM. A slowly varying field is effectively corrected by $(\partial_y^2 E_x)g^{yy}_{\mathbf{k}}\sim q^2 g^{yy}_{\mathbf{k}}E_x$, and the anomalous response to this effective field is governed by the Berry curvature in the charge Hall case, while it is governed by the orbital Berry curvature in the present OAM response.

In the second contribution in Eq.~\eqref{Eq_Result_ours_1}, we define the quantities $T^{L_{z}}_{yyx,\mathbf{k}}$ and $T^{L{z}}_{yyy,\mathbf{k}}$, which involve three spatial indices, as in Eq.~\eqref{Eq_T_yyx}. These are introduced in analogy with the symmetrized tensor of the quantum geometric connection in Eq.~\eqref{Eq_Tijl_1}, which appears in the charge conductivity~\cite{kozii2021intrinsic}. Although our quantities are not tensors in a strict sense, both their index structure and coefficients involving the velocity difference correspond to those of the symmetrized tensor in the charge conductivity reported previously~\cite{kozii2021intrinsic,note_XY_Axes}. Another term in the second contribution has the same derivative structure, with the Berry curvature replaced by the orbital Berry curvature~\cite{kozii2021intrinsic}.

Although the present analysis focuses on the local OAM contribution, possible wave-packet self-rotation effects can be incorporated within the modern theory of orbital magnetism~\cite{thonhauser2005orbital,vanderbilt_2018,lee2026anatomy}. In the present real two-band system, this self-rotation correction vanishes under the assumed symmetries and therefore does not generate an additional contribution to Eq.~\eqref{Eq_Result_ours_1}. The finite response obtained here is thus governed by the local OAM matrix element in Eq.~\eqref{Loperator}.

\section{Numerical Analysis}\label{sec_numerics}
We consider an anisotropic two-orbital model on a square lattice and compare the orientation dependence of the zeroth- and second-order orbital Hall responses. The zeroth-order response is governed by the orbital Berry curvature, whose mixed $k_x$- and $k_y$-derivative structure involves each momentum direction once. By contrast, the second-order conductivity $\sigma_{yx}^{L_z,(2)}$ contains additional momentum derivatives associated with the nonuniform field. The second-order response is therefore naturally expected to be more sensitive to the directional anisotropy of the band structure.

We use a minimal tight-binding model written in a real two-orbital basis, which may be represented by $(p_x,p_y)$ orbitals or, more broadly, by an effective $(d_{xz},d_{yz})$ two-orbital subspace~\cite{raghu2008minimal,kotani2007intrinsic}. The Hamiltonian density with nearest- and next-nearest-neighbor hoppings is written as
\begin{equation}
H(\mathbf{k})=
\begin{pmatrix}
\epsilon_x(\mathbf{k}) & \gamma(\mathbf{k})\\
\gamma(\mathbf{k}) & \epsilon_y(\mathbf{k})
\end{pmatrix},
\end{equation}
where
\begin{equation}
\begin{aligned}
\epsilon_x(\mathbf{k})=2t_x\cos k_x-2t_p\cos k_y,\\
\epsilon_y(\mathbf{k})=2t_y\cos k_y-2t_p\cos k_x,
\end{aligned}
\end{equation}
and $\gamma(\mathbf{k})=\lambda \sin k_x\sin k_y$~\cite{han2023microscopic}. Here, $t_x$ and $t_y$ characterize the direction-selective nearest-neighbor hoppings, whereas $t_p$ denotes the weaker nearest-neighbor hoppings in the orthogonal directions. In the $(p_x,p_y)$ interpretation, this direction selectivity naturally reflects the anisotropic orbital overlap. The parameter $\lambda$ controls the diagonal next-nearest-neighbor hybridization between the two orbitals, generating a nontrivial orbital texture and hence a finite orbital Berry curvature. The in-plane anisotropy is introduced by taking $t_x\neq t_y$. The same Hamiltonian may also be viewed as an effective $(d_{xz},d_{yz})$ model, since the orbital Hall response considered here is governed by the inter-orbital dynamics in a real two-orbital subspace rather than by the microscopic orbital label itself. This broader interpretation is natural because OHE has also been discussed in multi-orbital $d$-electron systems~\cite{tanaka2008intrinsic,kotani2009giant,jo2018gigantic,canonico2020orbital}.

To compare their orientation dependence, we consider two configurations related by a $90^\circ$ sample rotation while keeping the external-field geometry fixed. In practice, the rotated configuration is implemented by interchanging $t_x$ and $t_y$. For the numerical results shown below, we choose $t_x=1.05t$, $t_y=0.95t$, $t_p=0.2t$, and $\lambda=0.6t$ before rotation, where $t$ is used as the reference hopping scale.

\begin{figure}[t!]
    \centering
    \includegraphics[width=1.0\linewidth]{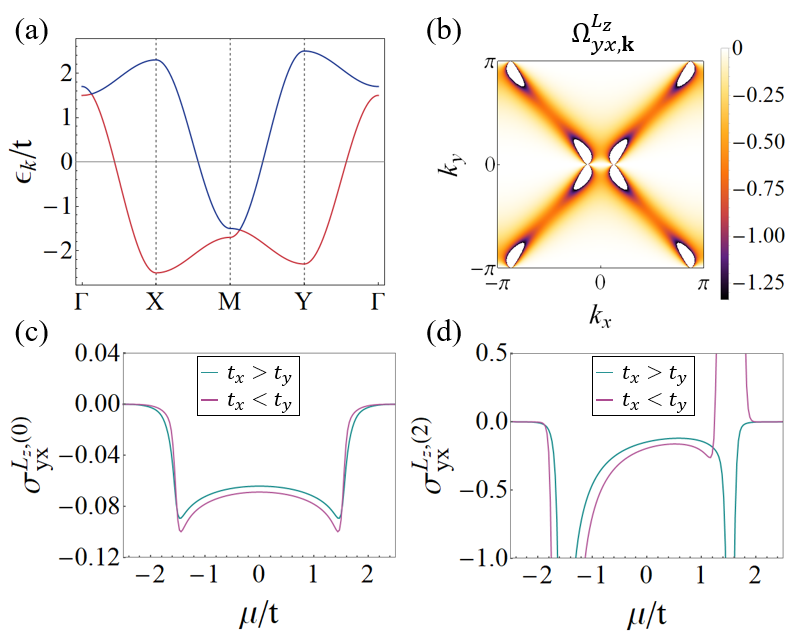}
    \caption{(a) Band structure of the model along the high-symmetry lines in the Brillouin zone and (b) orbital Berry curvature $\Omega^{L_z}_{yx,\mathbf{k}}$ in momentum space for the configuration with $t_x>t_y$. (c) $\sigma^{L_z,(0)}_{yx}$ and (d) $\sigma^{L_z,(2)}_{yx}$ as functions of the chemical potential $\mu$. In each of panels (c) and (d), the two curves represent two configurations related by interchanging $t_x$ and $t_y$, corresponding to a $90^\circ$ rotation of the anisotropic lattice while keeping the external-field geometry fixed. The energy and chemical potential are expressed in units of the reference hopping scale $t$. Common overall prefactors are omitted for the orbital Berry curvature and conductivities.}
    \label{fig2}
\end{figure}

Figure~\ref{fig2} summarizes the numerical results. The band structure and orbital Berry curvature in Figs.~\ref{fig2}(a) and \ref{fig2}(b) visualize the anisotropy of the model in momentum space. Figures~\ref{fig2}(c) and \ref{fig2}(d) then show how this anisotropy affects the transport response by comparing the chemical-potential dependence of $\sigma_{yx}^{L_z,(0)}$ and $\sigma_{yx}^{L_z,(2)}$ before and after the $90^\circ$ sample rotation. The comparison shows that $\sigma_{yx}^{L_z,(2)}$ exhibits a much stronger orientation dependence than $\sigma_{yx}^{L_z,(0)}$. 

To quantify the rotation-induced change, we evaluate the relative difference between the two orientations as
\begin{equation}
  \frac{\int_{\mathcal{D}}d\mu\,|\sigma_n^A(\mu)-\sigma_n^B(\mu)|}{\int_{\mathcal{D}}d\mu \, \frac{1}{2}\left(|\sigma_n^A(\mu)|+|\sigma_n^B(\mu)|\right)}.  
\end{equation}
Here, $A$ and $B$ denote the responses before and after the $90^\circ$ sample rotation, respectively. The domain $\mathcal{D}$ excludes a narrow region near the accidental band degeneracy, where the direct use of the nondegenerate Kubo formula and the subsequent small-$q$ Taylor expansion require extra care, as reflected in the conditions $q|A^y_{\mathbf{k}}|\ll 1$ and $|q\delta v^{y}_{\bf k}/\delta\epsilon_{\mathbf{k}}|\ll 1$.

The resulting relative difference is about $9.66\%$ for $\sigma_{yx}^{L_z,(0)}$, whereas it reaches about $52.7\%$ for $\sigma_{yx}^{L_z,(2)}$. This quantitative comparison confirms that the second-order orbital Hall response exhibits a much stronger orientation dependence in an anisotropic sample than the zeroth-order response, consistent with the additional momentum derivatives along the nonuniform-field direction contained in $\sigma_{yx}^{L_z,(2)}$. This enhanced orientation dependence may therefore provide a diagnostic feature for distinguishing the higher-order orbital Hall response from the conventional zeroth-order contribution in experiments.

\section{Conclusion}\label{Sec_conclusion}
We have investigated the OHE under a spatially nonuniform electric field by calculating the leading $q^2$-order correction to the orbital Hall conductivity. Our analysis assumes a two-dimensional electronic system that preserves both time-reversal and inversion symmetries. The resulting correction consists of five distinct terms, each of which has a direct counterpart in the second-order charge Hall conductivity reported in Ref.~\cite{kozii2021intrinsic}. In particular, our result can be obtained from the charge Hall conductivity by replacing the Berry curvature with the orbital Berry curvature. The quantum metric enters with the same form, and the remaining terms preserve the structure of their counterparts in the charge Hall conductivity. We further illustrate numerically, using a minimal two-orbital tight-binding model, that the higher-order response tends to be more sensitive to the orientation of an anisotropic sample than the conventional zeroth-order response. Our results highlight the role of quantum geometry in intrinsic orbital Hall transport.

\acknowledgements
This work was supported by the National Research Foundation of Korea (NRF) under Grant No. RS-2024-00410027. M.J.P. thanks the University of Alberta for its institutional support. J.M.L. acknowledges support from Quantum Horizons Alberta. We thank Jeonghun Sohn and Hojun Lee for discussions.

\appendix
\section{Orbital angular momentum operator}\label{Appx_OAM_Operator}
We discuss the OAM operator in Eq.~\eqref{Loperator}. For a real two-orbital basis, spinless time-reversal symmetry is represented by complex conjugation, $\mathcal{T}=\mathcal{K}$. Since the OAM operator is odd under time reversal, it satisfies
\begin{equation}
\mathcal{T} \hat{L}_{z} \mathcal{T}^{-1}=-\hat{L}_{z}.
\end{equation}
In the matrix form of the real-orbital basis, this condition gives
\begin{equation}
L_z^*=-L_z,
\end{equation}
showing that $L_z$ is purely imaginary. On the other hand, since $\hat{L}_z$ is an observable, its matrix representation must be Hermitian,
\begin{equation}
L_z^\dagger=L_z .
\end{equation}
The two properties above imply
\begin{equation}
L_z^T=-L_z,
\end{equation}
which means that $L_z$ is antisymmetric. Therefore, in the real-orbital basis, the $z$-component of the OAM operator is proportional to the Pauli matrix $\sigma_y$, with vanishing diagonal elements and oppositely signed off-diagonal elements.

\section{Derivation of the Kubo-Greenwood formula}\label{Appx_Kubo_1}
We consider the total system described by the Hamiltonian
\begin{equation}
    H = \sum_{\mathbf{k}} \left( \mathcal{H}^{0}_{\mathbf{k}} + \mathcal{H}'_{\mathbf{k}} \right),
\end{equation}
where $\mathbf{k}$ denotes the crystal momentum. 
Here, $\mathcal{H}^{0}_{\mathbf{k}}$ represents the periodic electronic Hamiltonian, and $\mathcal{H}'_{\mathbf{k}}$ describes the perturbative coupling between electrons and the external electric field. 
The eigenstates of the unperturbed system satisfy
\begin{equation}
    \mathcal{H}^{0}_{\mathbf{k}} |n_{\mathbf{k}}\rangle = \epsilon_{n,\mathbf{k}} |n_{\mathbf{k}}\rangle,
\end{equation}
where $|n_{\mathbf{k}}\rangle$ is the Bloch state of the $n$-th band and $\epsilon_{n,\mathbf{k}}$ is its corresponding band energy.

We apply an electric field along the $x$ direction,
\begin{equation}
    \mathbf{E} = \hat{x} E_{0} e^{i(qy - \omega t)} + \text{c.c.},
\end{equation}
which varies slowly in real space with a small wave vector $q$ along the $y$ direction. 
The perturbation term is then expressed as
\begin{equation}
    \mathcal{H}'_{\mathbf{k}} = \frac{E_{0}}{i\omega} \hat{j}^{x}_{\mathbf{k}} e^{i(qy - \omega t)} + \text{H.c.},
\end{equation}
where $\hat{j}^{x}_{\mathbf{k}} = e \hat{v}^{x}_{\mathbf{k}}$ denotes the charge current operator in the $x$ direction.

The perturbed state $|\tilde{n}_{\mathbf{k}}\rangle$ can be obtained using first-order perturbation theory as
\begin{equation}
\begin{aligned}
    |\tilde{n}_{\mathbf{k}}\rangle 
    = &\, |n_{\mathbf{k}}\rangle
    + \frac{E_{0}}{i\omega} 
    \sum_{m} 
    \Bigg[
        \frac{
        \langle m_{\mathbf{k}+q\hat{y}}|
        \hat{j}^{x}_{\mathbf{k}}
        |n_{\mathbf{k}}\rangle
        }{
        \epsilon_{n,\mathbf{k}} - \epsilon_{m,\mathbf{k}+q\hat{y}} + \omega
        }\\ 
        &\times |m_{\mathbf{k}+q\hat{y}}\rangle e^{i(qy - \omega t)}  
    - (q, \omega \!\to\! -q, -\omega)
    \Bigg],
\end{aligned}
\end{equation}
where $m,n$ denote band indices~\cite{marder2010condensed}.

Using the perturbed states, we calculate the OAM current. 
The operator for the OAM current is defined as
\begin{equation}
    \hat{j}^{L_{z}}_{y,\mathbf{k}} = 
    \frac{1}{2}
    \left( 
        \hat{L}_{z}\hat{v}^{y}_{\mathbf{k}} + 
        \hat{v}^{y}_{\mathbf{k}}\hat{L}_{z} 
    \right),
\end{equation}
where $\hat{v}^{y}_{\mathbf{k}}$ is the velocity operator in the $y$ direction, and $\hat{L}_{z}$ is the OAM operator along the $z$ axis~\cite{go2018intrinsic}.

The expectation value of the OAM current, calculated to first order in the perturbation, is given by
\begin{equation}
\begin{aligned}
    j^{L_{z}}_{y} 
    \simeq & \frac{1}{S}\sum_{n,\mathbf{k}} 
    f_{n,\mathbf{k}} 
    \langle \tilde{n}_{\mathbf{k}}| 
    \hat{j}^{L_{z}}_{y,\mathbf{k}} 
    | \tilde{n}_{\mathbf{k}}\rangle \\
    =&\, \frac{E_{0}}{i\omega S} 
    \sum_{n,m,\mathbf{k}} f_{n,\mathbf{k}}
    \Bigg[
        \frac{
        \langle n_{\mathbf{k}}| 
        \hat{j}^{L_{z}}_{y,\mathbf{k}} 
        |m_{\mathbf{k}+q\hat{y}}\rangle 
        \langle m_{\mathbf{k}+q\hat{y}}| 
        \hat{j}^{x}_{\mathbf{k}} 
        |n_{\mathbf{k}}\rangle
        }{   \epsilon_{n,\mathbf{k}} - \epsilon_{m,\mathbf{k}+q\hat{y}} + \omega }\\
        &+
        \frac{
        \langle n_{\mathbf{k}-q\hat{y}}| 
        \hat{j}^{L_{z}}_{y,\mathbf{k}} 
        |m_{\mathbf{k}}\rangle 
        \langle m_{\mathbf{k}}| 
        \hat{j}^{x}_{\mathbf{k}} 
        |n_{\mathbf{k}-q\hat{y}}\rangle
        }{
        \epsilon_{n,\mathbf{k}} - \epsilon_{m,\mathbf{k}-q\hat{y}} - \omega
        }
    \Bigg]e^{i(qy - \omega t)}\\
    &+ (q,\omega \!\to\! -q, \omega),
\end{aligned}
\end{equation}
where $S$ is the area of the system.

After some algebraic manipulation, the orbital Hall conductivity can be expressed as
\begin{equation}
\begin{aligned}
&\sigma^{L_{z}}_{yx}(q,\omega) = 
\sum_{n,m,\mathbf{k}} 
\frac{1}{i\omega S}
(f_{n,\mathbf{k}-q\hat{y}/2} - f_{m,\mathbf{k}+q\hat{y}/2})\\
&\times \frac{ \langle n_{\mathbf{k}-q\hat{y}/2}| 
\hat{j}^{L_{z}}_{y,\mathbf{k}} 
|m_{\mathbf{k}+q\hat{y}/2}\rangle 
\langle m_{\mathbf{k}+q\hat{y}/2}| 
\hat{j}^{x}_{\mathbf{k}} |n_{\mathbf{k}-q\hat{y}/2}\rangle }{\epsilon_{n,\mathbf{k}-q\hat{y}/2} - \epsilon_{m,\mathbf{k}+q\hat{y}/2} + \omega }.
\end{aligned}
\end{equation}
This expression follows from the definition
\begin{equation}
    j^{L_{z}}_{y} = 
    %\sum_{q,\omega} 
    \sigma^{L_{z}}_{yx}(q,\omega) 
    E_{0} e^{i(qy - \omega t)}+\text{c.c}.
\end{equation}
Here, the current matrix elements between the states at $\mathbf{k}-\mathbf{q}/2$ and $\mathbf{k}+\mathbf{q}/2$ are expressed using the current operator evaluated at the midpoint momentum $\mathbf{k}$. This step is exact for a quadratic continuum band and gives the leading small-$q$ form for a general Bloch band. Then, we perform the Wick rotation and take the dc limit $\omega \rightarrow 0$. By replacing the charge current operators with the velocity operators, we finally obtain the Kubo-Greenwood formula for the orbital Hall conductivity used in the main text~\cite{kozii2021intrinsic,zhang2022geometric}.

\bibliography{BibRef}

\end{document}